\documentclass[nofootinbib,aps,prstab,twocolumn,groupedaddress,showpacs,showkeys,superscriptaddress,reprint]{revtex4-1}
\usepackage[english]{babel}
\usepackage[utf8x]{inputenc}
\usepackage{amssymb}
\usepackage{amsthm}
\usepackage{amsmath}
\usepackage{hyperref}
\usepackage{subfigure}
\usepackage[percent]{overpic} 
\usepackage{picture}
\usepackage{color}
\usepackage{nth}
\usepackage{gensymb}
\usepackage{cleveref}
\usepackage{textcomp}
\usepackage{makecell}
\usepackage{lineno} 
\usepackage{textcomp}
\graphicspath{{img/}}

\begin{document}

\title{Plasma lens-based beam extraction and removal system for Plasma Wakefield Acceleration experiments}


\author{R. Pompili}
\email[]{riccardo.pompili@lnf.infn.it}
\author{E. Chiadroni}
\affiliation{Laboratori Nazionali di Frascati, Via Enrico Fermi 40, 00044 Frascati, Italy}
\author{A. Cianchi}
\affiliation{INFN and Tor Vergata University, Via Ricerca Scientifica 1, 00133 Rome, Italy}
\author{A. Del Dotto}
\affiliation{Laboratori Nazionali di Frascati, Via Enrico Fermi 40, 00044 Frascati, Italy}
\author{L. Faillace}
\affiliation{INFN Milano, via Celoria 16, 20133 Milan, Italy}
\author{M. Ferrario}
\affiliation{Laboratori Nazionali di Frascati, Via Enrico Fermi 40, 00044 Frascati, Italy}
\author{P. Iovine}
\author{M.R. Masullo}
\affiliation{INFN Napoli, Via Cintia, 80126 Naples, Italy}


\date{\today}

\begin{abstract}

Plasma Wakefield Acceleration represents one of the most promising techniques able to overcome the limits of conventional RF technology and make possible the development of compact accelerators. With respect to the laser-driven schemes, the beam-driven scenario is not limited by diffraction and dephasing issues, thus it allows to achieve larger acceleration lengths. One of the most prominent drawback, conversely, occurs at the end of the acceleration process and consists of removing the depleted high-charge driver while preserving the main features (emittance and peak current) of the accelerated witness bunch. Here we present a theoretical study demonstrating the possibility to reach these goals by using an innovative system consisting of an array of beam collimators and discharge-capillaries operating as active-plasma lenses. Such a system allows to extract and transport the accelerated and highly divergent witness bunch and, at the same time, provides for the removal of the driver. The study is completed by a set of numerical simulations conducted for different beam configurations. The physics of the interaction of particles with collimator is also investigated.

\end{abstract}


\maketitle

\section{Introduction}
Plasma-based acceleration, either driven by ultra-short laser pulses~\cite{1979PhRvL..43..267T,leemans2006gev,faure2006controlled} or electron bunches~\cite{2007Natur.445..741B,lu2009high,litos2014high}, represents one of the most promising techniques able to overcome the limits of conventional RF technology and allow the development of compact accelerators. In both schemes the plasma is used as an energy transformer in which the \textit{driver} pulse energy is transferred to the plasma (through the excitation of plasma wakes) and, in turn, to a \textit{witness} bunch (externally~\cite{rossiplasma,steinke2016multistage} or self-injected~\cite{faure2004laser,2004Natur.431..538G}).
Plasma Wakefield Acceleration (PWFA) driven by one or more \textit{driver} bunches offers several advantages with respect to laser-driven schemes, mainly limited in their overall efficiency and in the maximum accelerating lengths that can be achieved considering the laser diffraction, dephasing and depletion~\cite{1996ITPS...24..252E}. 
At FACET~\cite{litos2014high}, for instance, it has been experimentally demonstrated that by employing driver beams with GeV energies the overall efficiency can be boosted up to $30\%$ on a 40~cm-long plasma stage.
It represents the starting point for future facilities based on plasma acceleration like, for instance, EuPRAXIA~\cite{walker2017horizon}.

A drawback common to both the laser and particle beam-driven methods is represented by the extraction of the accelerated witness bunch that can lead to a large degradation of its emittance. 
When exiting the plasma the accelerated bunch has a large angular divergence~\cite{dornmair2015emittance}, of several mrad, that is some orders of magnitude larger with respect to beams accelerated by conventional (RF) photo-injectors.
For non-negligible energy spreads, such a large divergence also leads to a rapid increase of the normalized emittance in the downstream drift~\cite{migliorati2013intrinsic}. It is thus mandatory to catch the accelerated witness as soon as possible.
In addition to this, for PWFA another issue is represented by the removal of the high-charge energy-depleted driver(s)~\cite{pompili2016beam}.
In this case it is evident that simple collimating apertures can not be employed without affecting also the witness charge.

To solve this problem, here we discuss about an innovative and compact witness extraction system that aims to solve these points by implementing an array of active-plasma lenses~\cite{van2015active,pompili2017experimental,chiadroni2018overview} consisting of multiple discharge-waveguide capillaries filled with gas. Taking advantage of the symmetric and linear ($\propto 1/\gamma$) focusing provided by an active-plasma lens~\cite{van2017nonuniform,pompili2018guiding}, such a solution allows to catch and transport the witness along the array while over-focusing the (lower energy) driver that is removed by using a collimator located between the lenses.
Our study indicates that the entire system can be made very compact (tens of centimetres scale) and adapted to different beam configurations (from hundreds MeV up to several GeV energies).

The paper is organized as follows. Sec.~\ref{eupraxia_sec} describes the reference scenario, i.e. a typical driver-witness beam configuration as the one foreseen for the proposed EuPRAXIA~\cite{walker2017horizon} design study. In sec.~\ref{the_system} we describe the conceptual design of the extraction system and discusses about the working regime of the employed plasma lenses. Beam dynamics simulations are then reported in sec.~\ref{beam_dynamics} where the scalability of the system is tested for several beam configurations. 
In sec.~\ref{collimator_sec} we describe the interactions of the particle beams with the collimator device. The analysis highlights the effects of the wakefields generated in the collimator aperture (sec.~\ref{wake_sec}) and the particle-matter interactions by means of the GEANT4 framework (sec.~\ref{geant_sim}), allowing to evaluate the effective removal of the driver particles and the effects of their interaction with the surrounding materials.

\section{The EuPRAXIA design study}\label{eupraxia_sec}
In the context of accelerator research, a fundamental milestone is represented by the realization of a plasma-driven facility that will integrate high-gradient accelerating plasma modules with a short-wavelength Free Electron Laser (FEL). In such a context, the Horizon 2020 design study EuPRAXIA (European Plasma Research Accelerator with eXcellence In Applications) project aims at designing the world's first accelerator facility based on advanced plasma-wakefield techniques to deliver 1-5~GeV electron beams that simultaneously have high charge, low emittance and low energy spread, which are required for applications by future user communities~\cite{walker2017horizon}.
The EuPRAXIA collaboration will conclude his work by suggesting a solid strategy with the aim to demonstrate the possibility to use plasma accelerators for user applications. The design study will propose a first European Research Infrastructure that is dedicated to demonstrate exploitation of plasma accelerators for users. Developing a consistent set of beam parameters produced by a plasma accelerator able to drive a short wavelength FEL is one of the major commitments of the design study.
To realize such a concept, EuPRAXIA foresees the realization of two facilities that will be the pillars of the design study. One will exploit the use of high-power lasers to generate the plasma wakes and provide the acceleration of an externally-injected witness bunch~\cite{pousa2019compact}. The other one will use beam-driven schemes, where the plasma wakefields are generated by one or more pre-accelerated electron bunches~\cite{ferrario2013sparc_lab}.

\begin{figure}[h]
\centering
\includegraphics[width=1.0\linewidth]{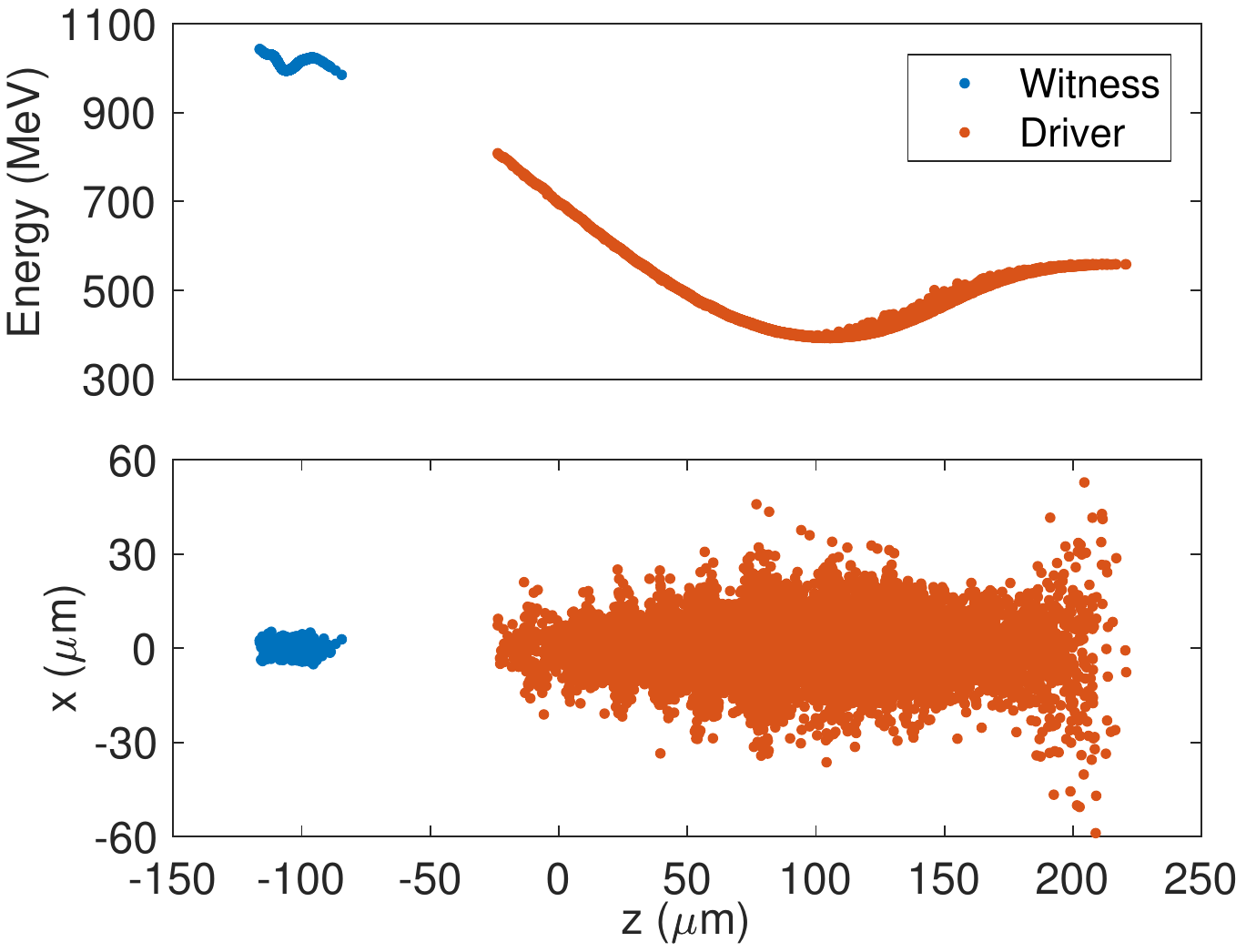}
\caption{Longitudinal phase-space (top) and \textit{x-z} view (bottom) of the driver (red) and witness (blue) bunches downstream the PWFA module. Most of the driver energy ($\approx 500$~MeV before the plasma interaction) has been depleted into the plasma and used to produce the accelerating field that boosted the witness energy up to $\approx 1$~GeV.}
\label{EU-LPS}
\end{figure}

For the beam-driven scenario, the LNF-INFN laboratories located in Frascati have been recently selected as possible hosting site~\cite{ferrario2018eupraxia,pompili_eupraxia}.
To generate FEL radiation, such a facility will make use of the X-band linac to produce and pre-accelerate a driver-witness beam up to 500~MeV and then inject it into a PWFA-based booster (operating with plasma densities of the order of $n_p\approx 10^{16}$~cm$^{-3}$) to increase the final energy up to 1~GeV~\cite{vaccarezza2018eupraxia,giribono2018eupraxia}.
Figure~\ref{EU-LPS} shows the simulated longitudinal phase-space (LPS) of a 200~pC driver followed by a 30~pC witness downstream the PWFA module. The results is obtained using Architect~\cite{marocchino2016efficient}, a hybrid code that works as a Particle-In-Cell (PIC) for the electron bunches while treating the plasma as a fluid. As shown, most of the driver energy has been depleted into the plasma and converted into plasma wakefields used to accelerate the witness bunch.
The simulation has been performed to demonstrate that it is possible to preserve the witness bunch quality (emittance and energy spread) during the acceleration in a plasma channel~\cite{pompili2016beam}. The main parameters of the two bunches at the exit of the PWFA module are reported in Tab.~\ref{tab_parameters}.

\begin{table}[ht]
\setlength{\tabcolsep}{6pt} 
\begin{center}
    \begin{tabular}{llcc}
	\hline
	\hline
	\textbf{Parameter}&\textbf{Units}&\textbf{Witness}&\textbf{Driver}\\	
	\hline
    Charge & pC & 30&	200\\
    Duration (rms)& fs& 11.5&		160\\
    Peak current& kA& 2.6&	1.2\\
    Energy & MeV & 1016&	460\\
    Energy Spread (rms)& $\%$& 0.73&	16\\
    Emittance& $\mu m$& 0.6&	5\\
    Spot size &	$\mu m$&	1.2&	7\\
	\hline
    \hline
    \end{tabular}
\end{center}
\caption{Witness and driver bunches parameters at the exit of the PWFA module.}
\label{tab_parameters}
\end{table}

\section{Extraction system}\label{the_system}
When exiting the plasma (with density $n_p$), the electron bunch moves from an extremely intense focusing field (generated by the excited plasma wakefield) to a free space, where the focusing effect suddenly vanishes.
The bunch is thus characterized by a large divergence $\sigma_{x'}\propto 1/\beta_{eq}$ at the plasma exit, where $\beta_{eq} = \sqrt{\gamma/{2\pi r_e n_p}}$ ($r_e$ is the classical electron radius and $\gamma$ the relativistic Lorentz factor) is
the witness Twiss $\beta$-function matched to the plasma~\cite{barov1994propagation}. As a reference case, when considering $n_p\approx 10^{16}$~cm$^{-3}$ and $\gamma\approx 2000$ (i.e. 1~GeV energy), one has $\beta_{eq}\approx 3$~mm. Thus, for typical normalized emittances of $\epsilon_n\approx 1~\mu m$, the angular divergence is of several mrad.
For non-negligible energy spreads $\sigma_E$, such a large divergence also leads to a rapid increase of the normalized emittance in a drift of length $s$ given by~\cite{migliorati2013intrinsic}
\begin{equation}
\label{emit_eq}
\epsilon_n^2 = \langle{\gamma}\rangle^2 \left({s^2 \left({\sigma_E \over E}\right)^2 \sigma_{x'}^4 + \epsilon_g^2}\right)~,
\end{equation}
where $E$ and $\epsilon_g$ are the bunch energy and geometrical emittance, respectively. According to eq.~\ref{emit_eq} it is thus important to catch the accelerated witness bunch with a short focal length focusing system installed downstream the plasma stage and as close as possible to it~\cite{pompili_pmq}.

\begin{figure}[h]
\centering
\includegraphics[width=1.0\linewidth]{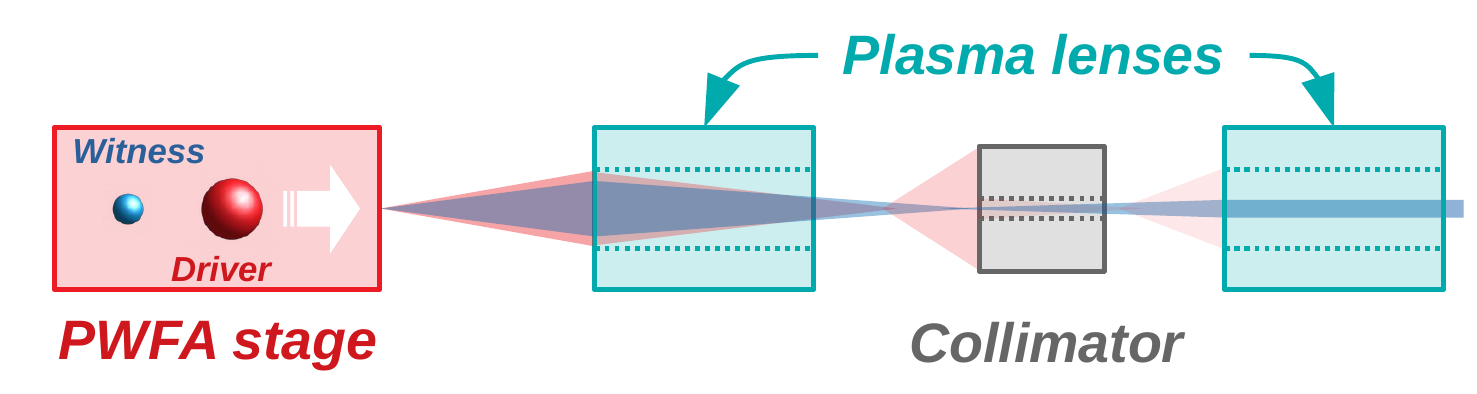}
\caption{Layout of the extraction system consisting of two active-plasma lenses employed as plasma lenses and a lead collimator between them. The focusing strength of the lenses are set up to rapidly catch the witness bunch (blue) downstream the PWFA stage and allow for its transport without any loss of charge. The driver bunch (red) has a lower average energy and is thus over-focused: its spot size at the collimator entrance is larger than the aperture and its charge is thus cut proportionally.}
\label{ExtractionSystem_sketch}
\end{figure}

The technique we are going to discuss has two goals: (i) provide an efficient capture of the witness bunch by preserving its emittance and peak current; (ii) remove the high-charge driver bunch during the transport in order to completely waste it before the FEL undulator beamline.
Figure~\ref{ExtractionSystem_sketch} shows the layout of such a system. The capture and focusing of the witness is provided by two active-plasma lenses (APL)~\cite{panofsky1950focusing}, being such devices able to produce large focusing fields of the order of kT/m~\cite{van2015active}. Between the APLs, a lead collimator is used to remove the driver bunch. The removal is based on the different focusing provided by the first APL to the two beams: the witness is focused exactly at the entrance of the collimator (so no charge is cut) while the driver (with approximately half of the energy) is over-focused to have a spot size larger than the collimator clear aperture.


To achieve the desired goals, i.e. capture of the high-current and low-emittance witness bunch and dump the high-charge and energy depleted driver, several simulation tools have been used and linked each other.
As described in sec.~\ref{eupraxia_sec}, the plasma acceleration process was simulated with the Architect code. To propagate and transport the two bunches from the exit of the PWFA module we have used the General Particle Tracer (GPT) code~\cite{de1996general}, a tool widely employed in the accelerator community. GPT has thus been used to simulate the drifts between all the elements involved in the system depicted in Fig.~\ref{ExtractionSystem_sketch} with space-charge effects included. In sec.~\ref{apl_dynamics} the dynamics of the beam into the plasma lenses is then computed by a 2D code written to solve the wakefield equations in the linear regime~\cite{fang2014effect}.
Finally, the optimization of the collimator system has been conducted to maximize the driver beam dumping by looking at its fundamental interactions with the collimator walls (predicted by GEANT4, see sec.~\ref{geant_sim}) and by analyzing the effects on the witness beam of the wakefields produced during the propagation through the collimator aperture (with the CST simulation framework, sec.~\ref{wake_sec}).

\subsection{Active-plasma lens}
\begin{figure}[h]
\centering
\includegraphics[width=1.0\linewidth]{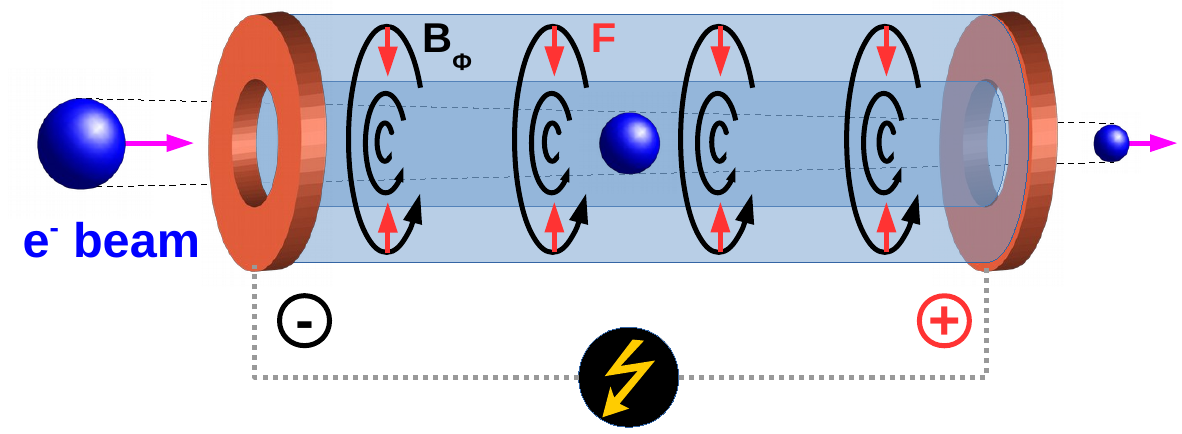}
\caption{Representation of the Active-Plasma Lens working mechanism. A discharge current is applied to the capillary through two symmetric electrodes. The current generates an azimuthal magnetic field ($\mathbf{B}_{\phi}$) that produces a focusing force ($\mathbf{F}$) for the incoming electron beam.}
\label{PlasmaLens_sketch}
\end{figure}

The core part of the system is represented by the active-plasma lenses.
An APL essentially consists of a current-carrying cylindrical conductor whose axis is parallel to the beam.
Here the plasma (produced after the ionization of the gas confined within the capillary) only acts as a conductor, while the net focusing effect is produced by the flowing discharge current.
A schematic picture of such a device is shown in Fig.~\ref{PlasmaLens_sketch}. By indicating with $J(r)$ the current density along the radial dimension, the resulting magnetic field is simply given by the Ampere law
\begin{equation}
\label{ampere_law}
B_{ext}(r) = {\mu_0\over r} \int_0^r J(r') r' dr'~, 
\end{equation}
where $\mu_0$ is the vacuum permeability. If the current density is uniform, the magnetic field increases linearly with the radius, and a linear restoring force on the beam will result.
One can notice three interesting features of such a device. (i) The focusing is symmetric, like in solenoids, and the focusing force scales as $F\propto \gamma^{-1}$ (with $\gamma$ the relativistic Lorentz factor), like in quadrupoles. 
(ii) The focusing can reach several tens of kT/m, i.e. orders of magnitude larger than the strongest available quadrupoles ($\approx 600$~T/m~\cite{lim2005adjustable,pompili2018compact}) and (iii) it is tunable by adjusting the external discharge current $I_D=\int_S \mathbf{J}\cdot d\mathbf{S}$.

These features pushed the research toward the use of APLs in accelerator facilities. While in past decades several results have been obtained by focusing ion beams~\cite{boggasch1991z,boggasch1992plasma,tauschwitz1996plasma}, recently several proof-of-principle experiments have been performed with laser-plasma~\cite{van2015active,van2017nonuniform} and RF~\cite{pompili2017experimental,lens_alberto,pompili2018focusing,lindstrom2018emittance} accelerators.
Although these results indicate the capability to integrate APLs in accelerator facilities and their advantages (in terms of focusing gradients) with respect to conventional optics, some of them have demonstrated that non-uniformities on the $J(r)$ and the interaction of the travelling beam with the background plasma can induce severe effects on the beam itself, in particular on its emittance~\cite{pompili2017experimental,lens_alberto}.

\begin{figure}[h]
\centering
\includegraphics[width=1.0\linewidth]{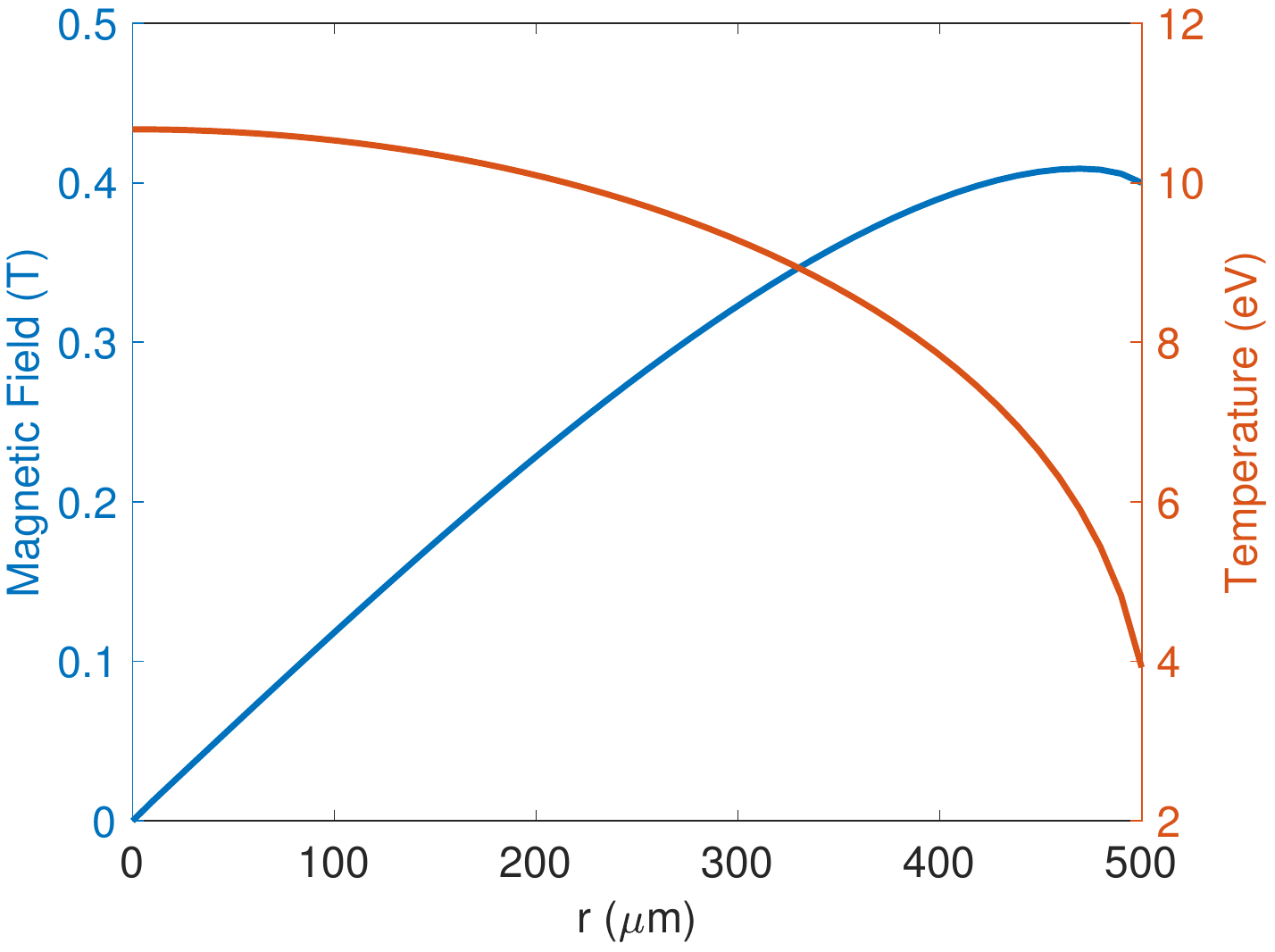}
\caption{Calculated profiles of the magnetic field (blue) and plasma temperature (red) along the capillary radius. The current-discharge used for the calculation is $I_D=1$~kA. The gas density is set to $n_{gas}=10^{16}$~cm$^{-3}$.}
\label{Bfield}
\end{figure}

The focusing field produced by the APL strongly depends on the discharge dynamics along the capillary. To describe the main effects of the discharge process, a one-dimensional analytical model can be used~\cite{bobrova2001simulations} by assuming the distribution of plasma inside the capillary at the equilibrium stage as soon as the discharge is initiated. In this case the equilibrium is determined only by the balance between Ohmic heating and cooling due to the electron heat conduction.
Figure~\ref{Bfield} shows the resulting magnetic field and plasma temperature computed for a capillary with $R_c=500~\mu m$ and a current discharge of $I_D=1$~kA. The initial density of the neutral H$_2$ gas is $n_{gas}=10^{16}$~cm$^{-3}$.
The nonlinearity of the magnetic field is one of the source leading to the emittance growth of the beam during focusing, although a recent work demonstrated the possibility to mitigate these effects~\cite{pompili2018focusing}.

\section{System performances}\label{beam_dynamics}
The extraction system we are discussing exploits the the different focusing provided by an APL on beams with different energies~\cite{van2018comparative}. This allows to tune the APL to transport the witness through the collimator and, at the same time, over-focus the driver and cut its charge.

\begin{figure}[h]
\centering
\includegraphics[width=1.0\linewidth]{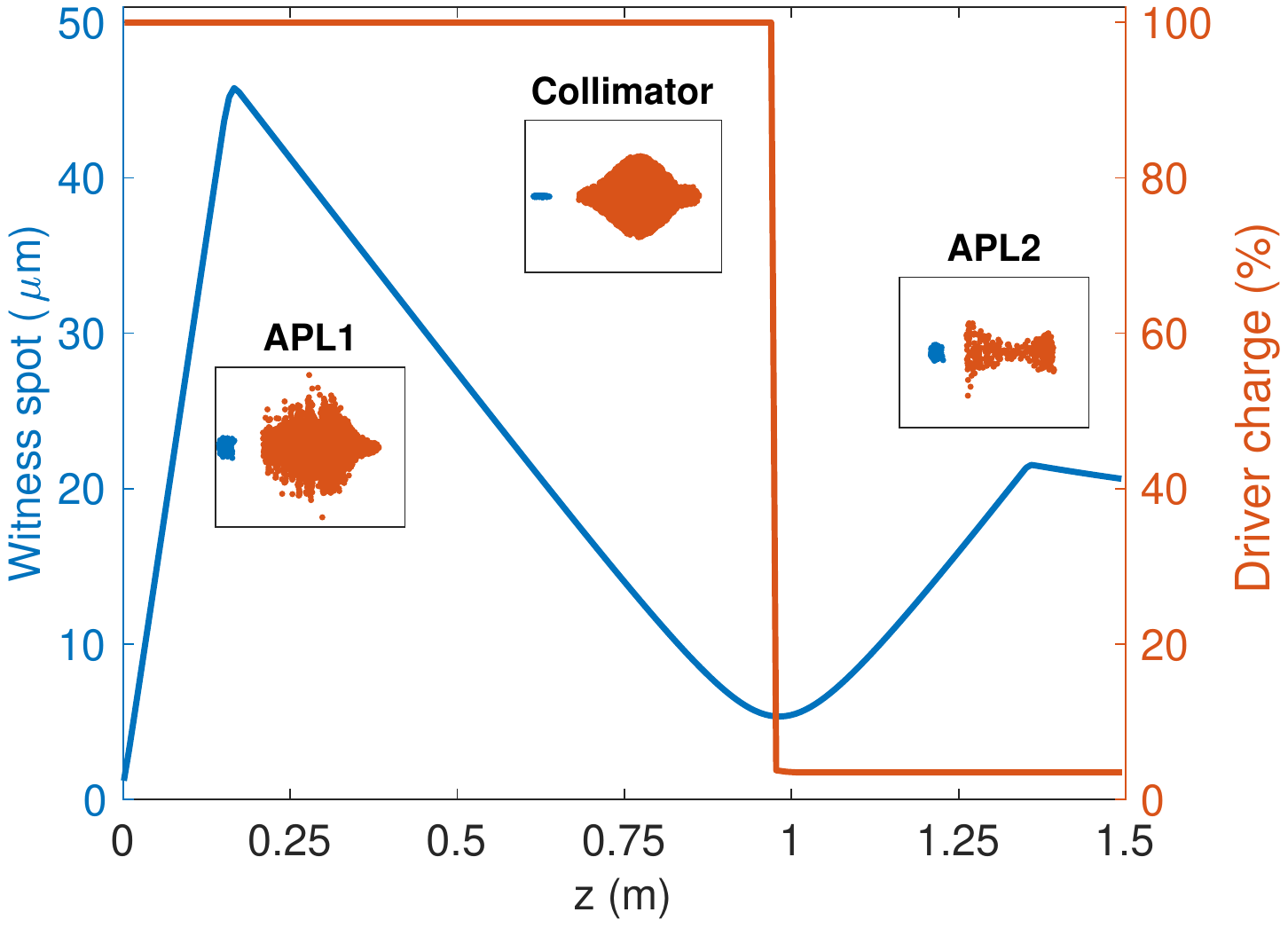}
\caption{Evolution of the witness envelope (blue) and driver charge (red) along the beamline. The insets show the \textit{x-z} plane of the bunches at the entrance of the \nth{1}~APL, the collimator and the \nth{2}~APL.}
\label{ChargeEvo}
\end{figure}

Figure~\ref{ChargeEvo} shows the evolution of the witness envelope (blue) and driver charge (red) along the system, consisting of two APLs and a lead collimator between them.
The system has been optimized on the LPS at the exit of the PWFA module, as the one reported in Fig.~\ref{EU-LPS}. The first lens consists of a 2~cm-long discharge-capillary with $500~\mu m$ hole radius. The focusing is obtained by applying a discharge current $I_D=1$~kA. The position of the lens with respect to the PWFA module has been carefully chosen to preserve as much as possible the witness emittance. On one side, according to eq.~\ref{emit_eq}, short drifts would be preferable to avoid emittance degradation due to the large divergence of the beam. On the other side, small drifts would produce a small witness spot at the APL entrance, i.e. a larger bunch density. As demonstrated in our previous work~\cite{pompili2018focusing}, large bunch densities produce non negligible plasma wakefields in the APL that, being nonlinear, would increase the beam emittance during the lens focusing. A trade-off is thus necessary to balance these two contributions. 
In Fig.~\ref{ChargeEvo} the best compromise has been found by moving the entrance on the first APL 15~cm downstream the PWFA module. In this case the normalized emittance increased in the drift from 0.6 to $0.7~\mu m$ (rms).

\begin{figure}[h]
\centering
\includegraphics[width=1.0\linewidth]{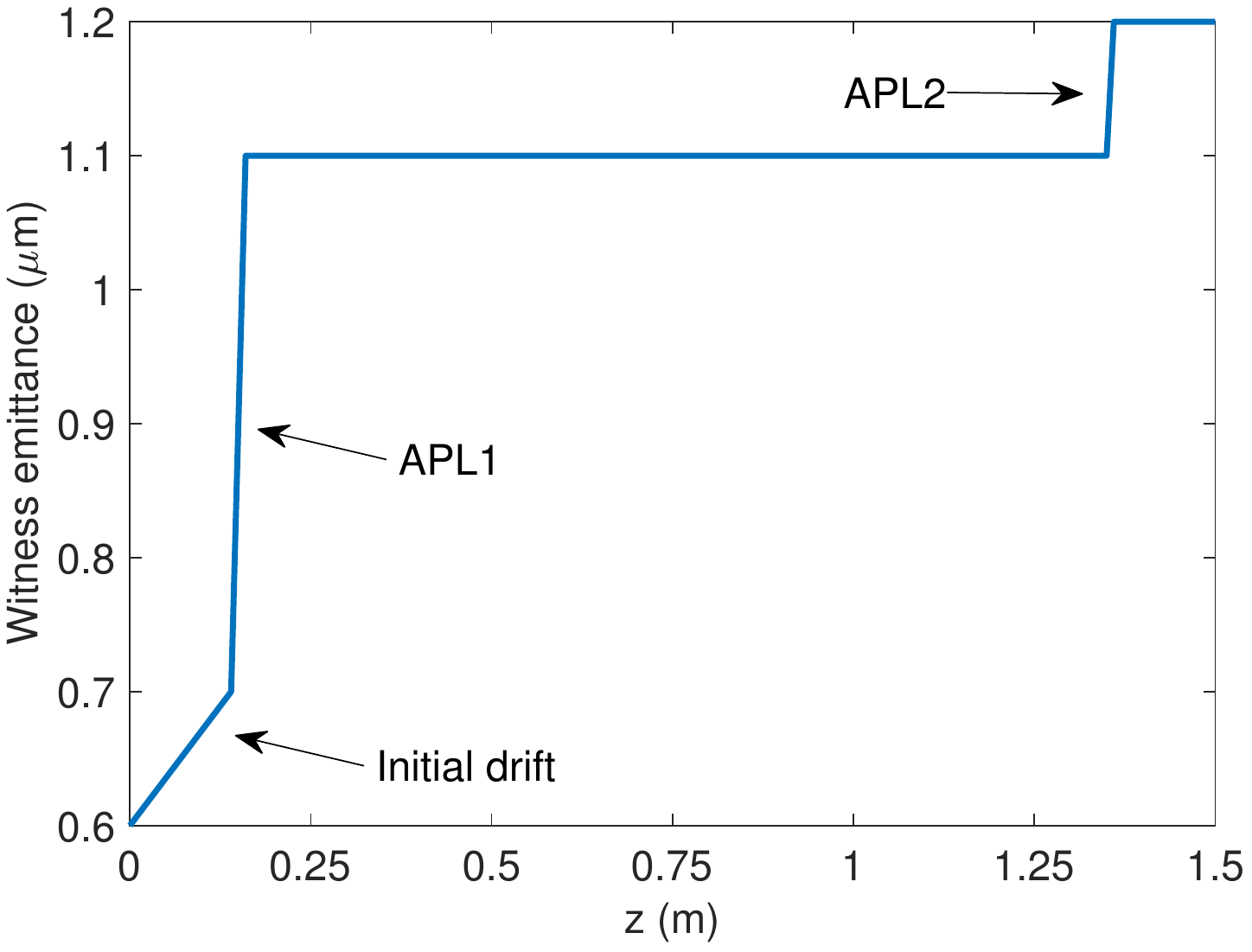}
\caption{Evolution of the witness emittance along the beamline. The points where there is an increase of such a quantity are labeled accordingly.}
\label{WitEmitEvolTot}
\end{figure}

The driver and witness are then focused up to the entrance of the collimator whose radius is $r_{coll}=200~\mu m$. Here the witness spot size is approximately $10~\mu m$ (rms), while the driver is almost twenty times larger.
In such a way the majority of the driver charge ($\approx 98\%$) is removed by the collimator, with only 4~pC that have remained after it; more details about the choice of the collimator aperture and particle interactions with its body are reported in sec.~\ref{geant_sim}. The system ends with the second APL (1~cm-long with $I_D=0.6$~kA) that catches the witness downstream the collimator and allow for its transport in the rest of the beamline.
Figure~\ref{WitEmitEvolTot} shows the evolution of the witness emittance along the beamline. As expected the increase of such a quantity is foreseen in the initial drift downstream the PWFA module (according to eq.~\ref{emit_eq}) and in the two active-plasma lenses (more details are reported in sec.~\ref{apl_dynamics}).
Considering all the elements and drift spaces involved, the total length of the system is 1.4~m.
All the details of the so developed beamline are summarized in Tab.~\ref{system_properties}.

\begin{table}[ht]
\setlength{\tabcolsep}{6pt} 
\begin{center}
    \begin{tabular}{lllll}
	\hline
	\hline
	\textbf{Element}&\textbf{Length}&\textbf{Radius}&\textbf{Position}&\textbf{Current}\\	
	\hline
	APL~1	&	2~cm	&	$500~\mu m$	&	15~cm	&	1~kA\\
	Collimator	&	3~cm	&	$200~\mu m$	&	97~cm	&	\\
	APL~2	&	1~cm	&	$500~\mu m$	&	135~cm	&	0.6~kA\\
	\hline
    \hline
    \end{tabular}
\end{center}
\caption{Optimized parameters for the APLs and collimator used in the proposed extraction system. The position of each element is relative to the exit of the PWFA module.}
\label{system_properties}
\end{table}

\subsection{Beam dynamics in the active-plasma lenses}\label{apl_dynamics}
As previously discussed, the focusing and guiding of an electron beam in an APL is actually due to the magnetic field generated by the discharge-current, while the plasma only acts as a conducting medium. However the dynamics of a beam is also affected by the interaction with the plasma that induce plasma wakefields acting on the beam itself~\cite{rosenzweig2004energy,lu2006nonlinear,shpakov2019longitudinal}.
In the limit of an electron beam with density comparable or smaller than the plasma one ($n_b\lesssim n_e$), the wakefields can be described by using the 2D plasma wakefield theory in linear regime~\cite{fang2014effect}.
If we rewrite the bunch density by separating the longitudinal and transverse profiles as $n_b(z,r)=n_{b,l}(z)\cdot n_{b,t}(r)$, the wakefields can be expressed as
\begin{equation}
\label{Wz_eq}
\begin{split}
W_z(z,r)=&{q_e\over \epsilon_0}R(r)\cdot\\ &\int_{-\infty}^z n_{b,l}(z')\cos \left({k_p(z-z')}\right)dz'\\
\end{split}
\end{equation}
\begin{equation}
\label{Wr_eq}
\begin{split}
W_r(s,r)=&{q_e\over {\epsilon_0 k_p}} {{dR(r)}\over dr} \cdot\\ &\int_{-\infty}^z n_{b,l}(z')\sin \left({k_p(z-z')}\right)dz'
\end{split}
\end{equation}
where $\epsilon_0$ is the vacuum permittivity, $q_e$ the electron charge. We have also introduced the plasma wave-number $k_p$ and the function $R(r)$ defined as
\begin{equation}
\begin{split}
R(r)&= k_p^2 K_0(k_p r) \int_0^r n_{b,r}(r')I_0 (k_p r') r' dr'\\ &+ k_p^2 I_0(k_p r) \int_r^{\infty} n_{b,r}(r')K_0 (k_p r') r' dr'~,
\end{split}
\end{equation}
where $I_0$ and $K_0$ are the modified Bessel functions of zeroth and first order.
From eq.~\ref{Wz_eq} and eq.~\ref{Wr_eq} we can thus compute the forces $F_{z,r}=q_e W_{z,r}$ acting on the beam. As a consequence we can see that, on the transverse plane, the plasma produced a net focusing. Indeed, when a relativistic electron bunch travels in a plasma the space-charge field of the electron beam is canceled by the plasma, thus the beam is pinched by its own magnetic field. Such a mechanism happens in so-called \textit{passive} plasma lenses, widely investigated in past years~\cite{su1990plasma,hairapetian1994experimental,nakanishi1991direct}.

\begin{figure}[h]
\centering
\includegraphics[width=1.0\linewidth]{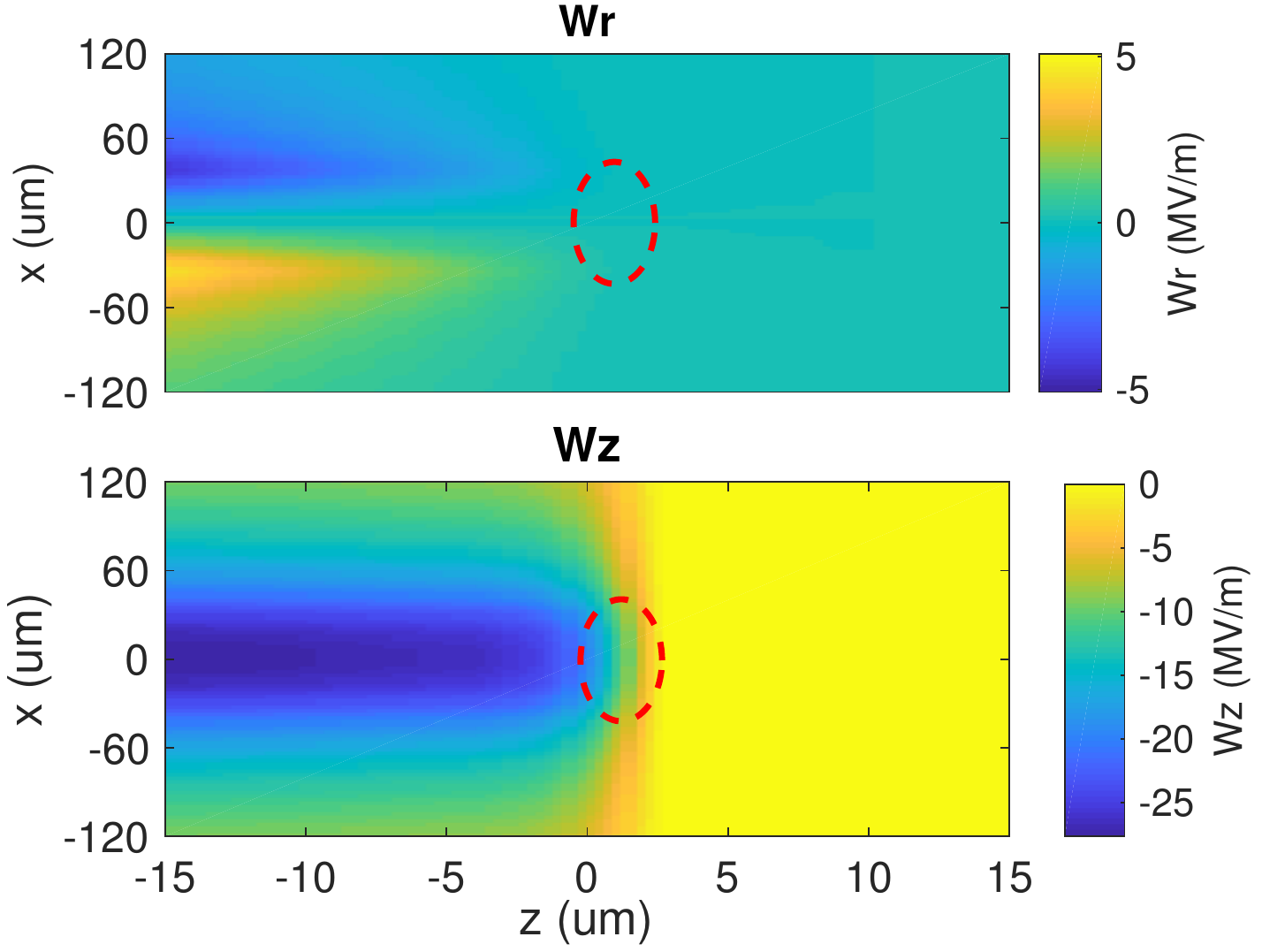}
\caption{Simulated radial (top) and longitudinal (bottom) plasma wakefield produced into the APL by the witness bunch. The red-dashed ellipse shows the effective position of the bunch.}
\label{WakeSnap}
\end{figure}

Starting from this set of equations we have developed a plasma simulation code that allows to track the evolution of a test electron beam.
Figure~\ref{WakeSnap} shows a snapshot of the radial ($W_r$) and longitudinal ($W_z$) wakefields induced by the 30~pC witness bunch moving in a plasma background with density $n_p=10^{16}$~cm$^{-3}$. We can see that transverse (longitudinal) fields as large as 5(30)~MV/m are produced. While the longitudinal dynamics is not affected, the transverse one is influenced both by the transverse plasma wakefield $W_r$ and the azimuthal magnetic field $B_{ext}$ induced by the discharge-current.

\begin{figure}[h]
\centering
\includegraphics[width=1.0\linewidth]{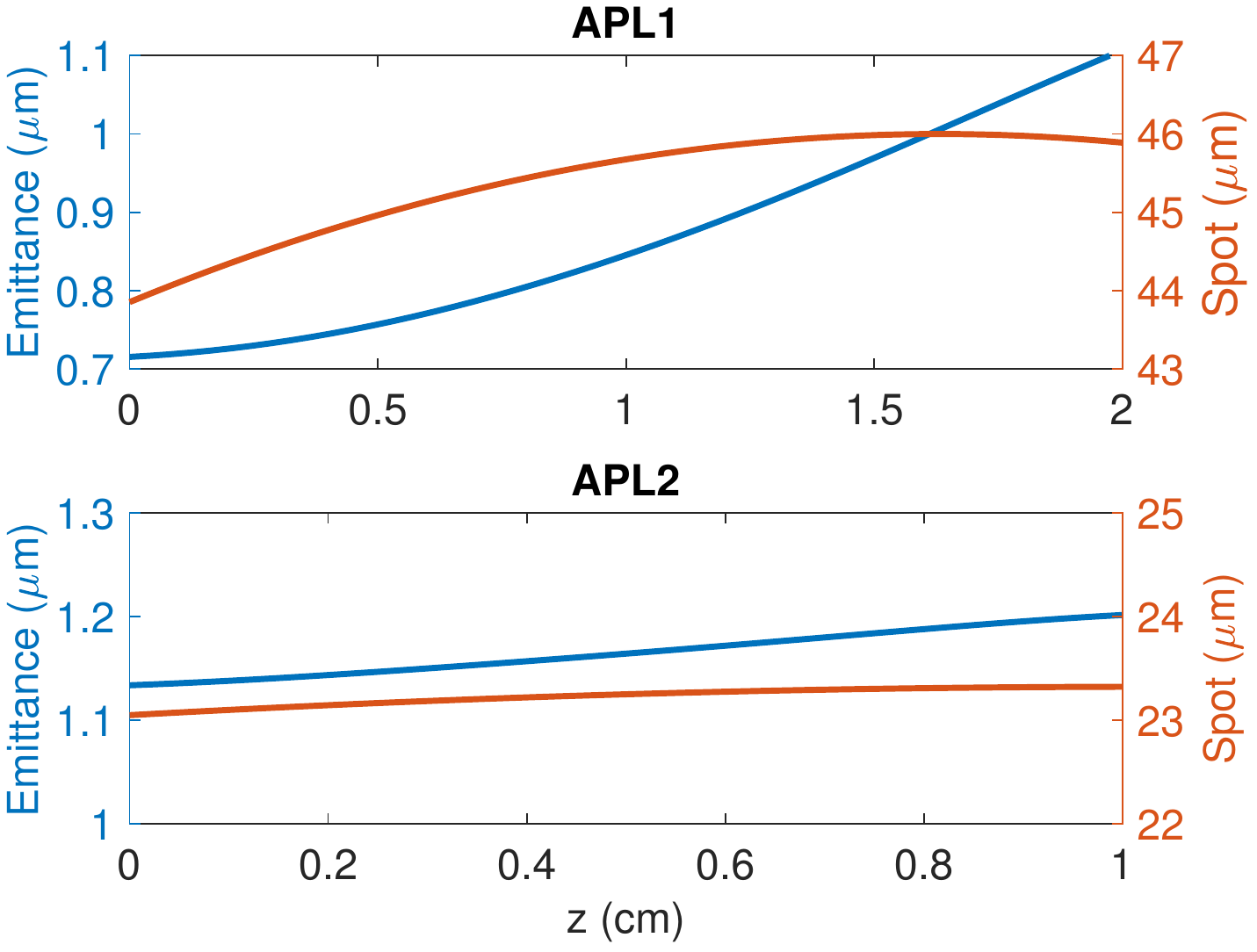}
\caption{Evolution of the witness emittance (blue) and spot size (red) along the two APLs.}
\label{APLevo}
\end{figure}

\begin{figure}[h]
\centering
\includegraphics[width=1.0\linewidth]{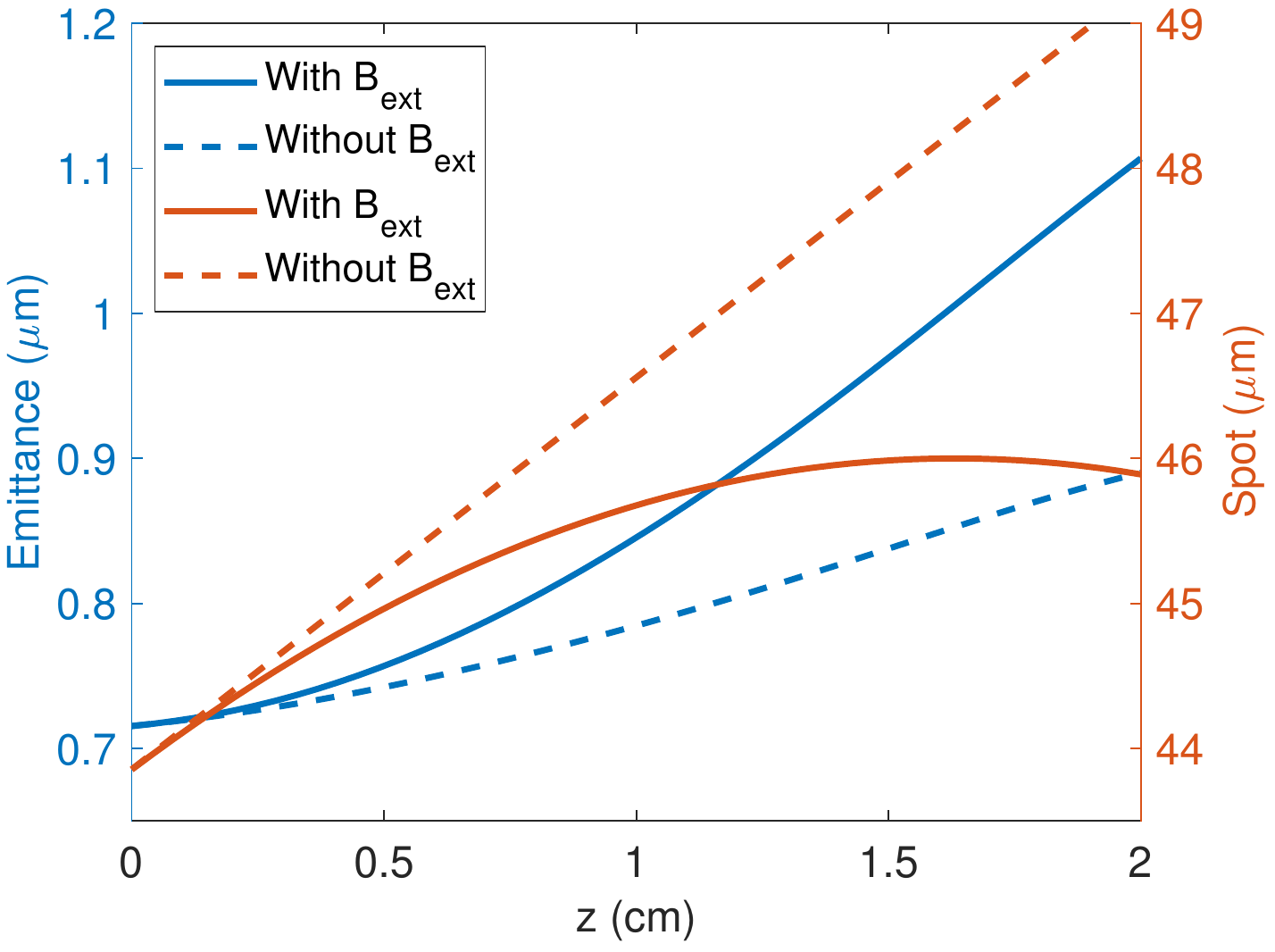}
\caption{Evolution of the witness emittance (blue) and spot size (red) along a 2~cm-long APL by including (solid) or not (dashed) the external azimuthal field produced by the discharge-current.}
\label{APLcomparison}
\end{figure}

Figure~\ref{APLevo} shows the evolution of the witness emittance and envelope along the two APLs used in the setup reported in Tab.~\ref{system_properties}.
As shown, most of the emittance growth happens in the first (and longer) lens, where the witness transverse spot size is larger ($43~\mu m$ rms) with respect to the one entering into the second lens ($23~\mu m$ rms).
Such a growth is due to both the nonlinearities of the $B_{ext}$ field (cf. Fig.\ref{Bfield}) and $W_r$.
Indeed we show in Fig.~\ref{APLcomparison} a comparison for the same two parameters, emittance and spot, with and without the external magnetic field in the first APL.
The increase of emittance due to the passive lensing (i.e. $W_r$) and the external field is approximately the same, i.e. $\Delta \epsilon_n\approx 0.2~\mu m$.
This scenario, however, represents the best compromise since smaller (larger) witness spot sizes would enhance the contribution due to the $W_r$ ($B_{ext}$) term and produce an overall larger emittance degradation.

\subsection{Parametric study}
The tunability of the system for different beam parameters is ensured by simply changing the discharge-current applied to the APLs in order to adjust the witness focusing and transport along the beamline. Once the transport optics is optimized on the witness side, the disposal of the driver charge mainly depends on its own 6D phase-space and can be maximized by properly tailoring the collimator geometry (aperture and length).
In this paragraph we discuss the dynamics of the driver bunch for different configurations while the witness bunch is assumed to be as in the reference case described so far. This allows us to not change the beamline optics and check how it reacts to the incoming driver.

\begin{figure}[h]
\centering
\subfigure{
\begin{overpic}[width=0.93\linewidth]{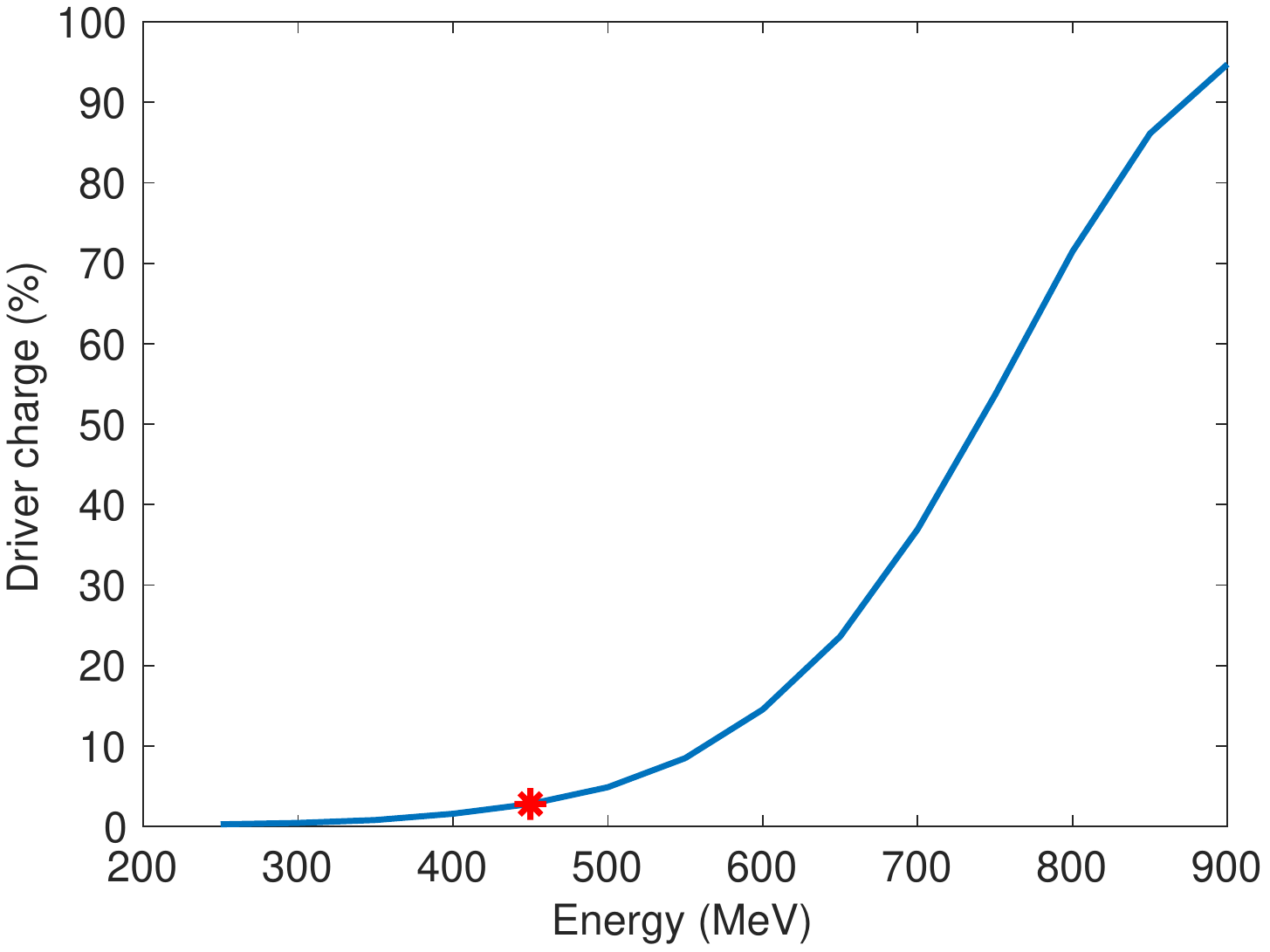}
\put(15,68){\color{black}\textbf{a}}
\end{overpic}
\label{dChargeEnergy_MR}
}
\subfigure{
\begin{overpic}[width=0.9\linewidth]{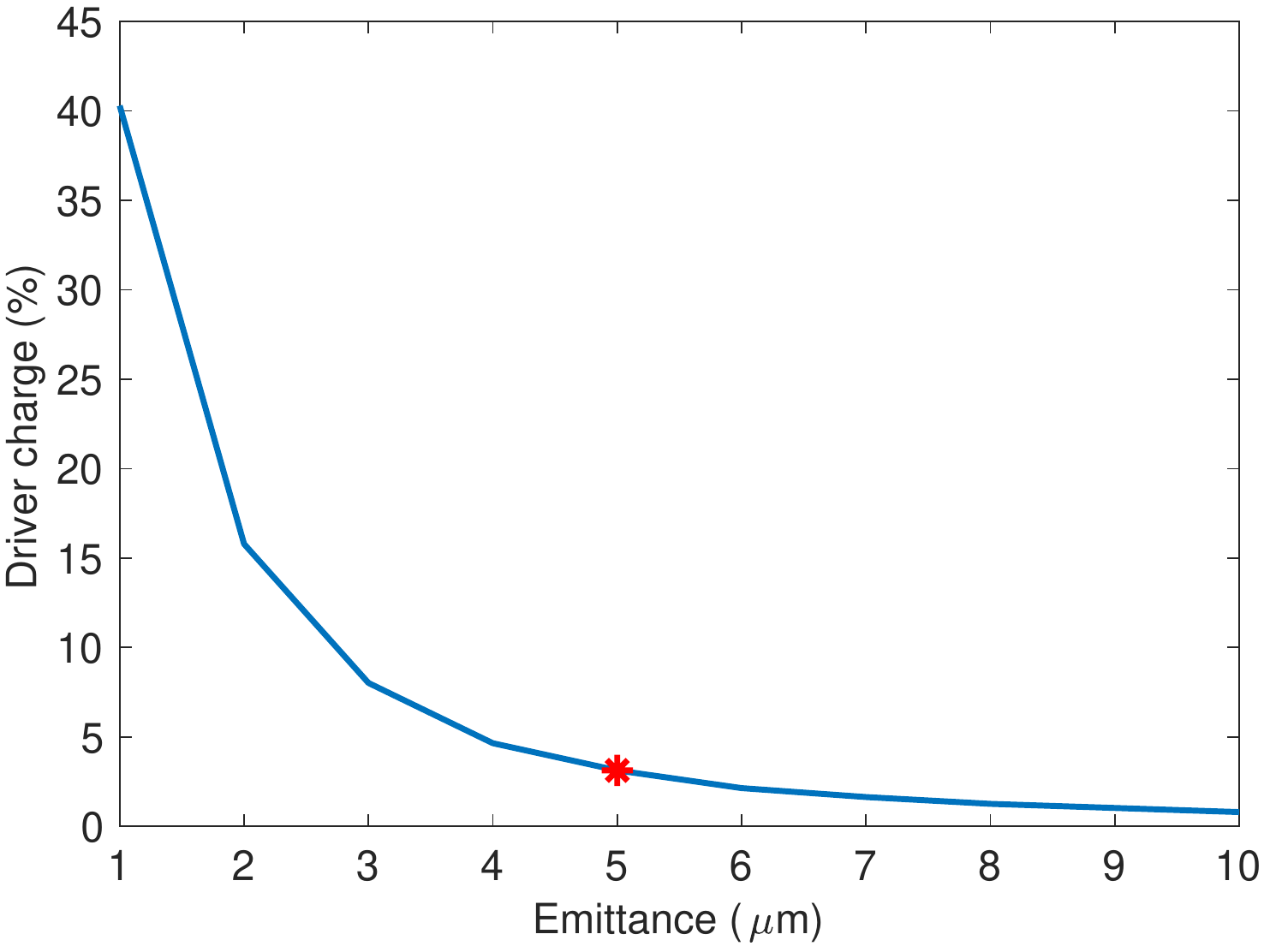}
\put(15,68){\color{black}\textbf{b}}
\end{overpic}
\label{dChargeEmittance_MR}
}
\caption{Driver charge at the end of the system for different driver energies (a) and normalized emittances (b). The red asterisks show the reference case described in Tab.~\ref{tab_parameters}.}
\label{MR_driver}
\end{figure}

Figure~\ref{dChargeEnergy_MR} shows how the charge-cut is affected according to the incoming beam energy. The system is tuned around the witness so we expect that for energies close to the witness one ($\approx 1$~GeV) the removal of the particles is less effective. We see that such a behaviour is basically obtained and lower is the beam energy, larger is the amount of charge cut by the system. On the contrary, at larger energies the system handles the beam as in the case of the witness, so its transport without charge-cut is achieved.
Another parameter that could change downstream the PWFA module is the bunch emittance, i.e. its divergence. Figure~\ref{dChargeEmittance_MR} shows how the final charge is dependent on such a parameter. As expected, larger beam divergences (meaning larger emittances) are favourable since the beam would reach the collimator with a larger transverse spot. On the contrary, low emittances mean that the beam is more collimated and the particles have almost parallel trajectories along the transport. This decrease the cut of charge operated by the collimator since less particles hit its walls.

In conclusion, the parametric study demonstrates that the system can be tuned on a particular configuration. For the sake of completeness we have also repeated this analysis by changing the driver spot size and energy spread, but these parameters only slightly affect the effective charge cut operated by the system (less than $10\%$ with respect to the reference case described so far) and have been omitted in this treatment.

\section{Beam interactions with the collimator}\label{collimator_sec}

\subsection{Wake potentials}\label{wake_sec}
The interaction between the bunch electric field and the conductive walls of the lead collimator described so far is able to generate wakefields that can affect both the longitudinal and transverse dynamics of the travelling bunch, hindering the emittance preservation~\cite{tenenbaum2002collimator}.
The electromagnetic interaction of charged particles with any surrounding environment can be quantified by solving the Maxwell equation to find the resulting electric and magnetic fields and then estimate the effects of these fields on the particle motion~\cite{palumbo1995wake,semyon1998impedances}.
Assuming that the \textit{z}-axis is the symmetry axis of the system under analysis and considering two particles that move along it, the electromagnetic field generated by a \textit{leading} charge $q_l$ (located at $z=ct$ with a transverse offset $r_1$ with respect to the \textit{z}-axis) produces a change of the momentum $\Delta \mathbf{p}$ on a \textit{trailing} particle with charge $q_t$ (located behind the first one at $s=ct-z$ with a transverse offset $r_2$).
Being the beam dynamics different in the longitudinal and transverse directions, it is useful to separate the longitudinal momentum $\Delta p_z$ from the transverse component $\Delta\mathbf{p}_{\perp}$ and introduce the longitudinal and transverse wake functions~\cite{semyon1998impedances} as
\begin{eqnarray}
w_z(r_1,r_2,s)&=&{{c\Delta p_z}\over{q_l q_t}}\\
\mathbf{w}_{\perp}(r_1,r_2,s)&=&{{c\Delta \mathbf{p}_{\perp}}\over{q_l q_t}}
\end{eqnarray}
Regarding the longitudinal wake, if the $r_{1,2}$ offsets are small in comparison to the aperture of the pipe we can remove the radial dependence and approximate $w_z(s)=w_z(0,0,s)$. For a bunch of longitudinal charge distribution $\rho(s)$, the longitudinal wake potential $W_z(s)$ (the voltage lost for a test particle at position $s$) is thus given by
\begin{equation}
\label{Wakez_eq}
W_z(s)=\int_0^{\infty} w(s') \rho(s-s') ds'~.
\end{equation}
In a similar way, by assuming $r_1=r_2=r$, the transverse wake potential $W_{\perp}(s)$ (that represents the transverse kick for a test particle at position \textit{s}) can be approximated to its lowest-order linear term ($\mathbf{w}_{\perp}(r,r,s)\approx \mathbf{w}_{\perp}(s)/r$) and expressed as
\begin{equation}
\label{Wperp_eq}
W_{\perp}(s)={1\over{r}}\int_0^{\infty} {|\mathbf{w}_{\perp}(s')|} \rho(s-s') ds'~.
\end{equation}
The average of the longitudinal wake potential from eq.~\ref{Wakez_eq} gives the loss factor $k_l$ while the average of the transverse wake from eq.~\ref{Wperp_eq} represents the kick factor $k_t$.

\begin{figure}[h]
\centering
\includegraphics[width=0.9\linewidth]{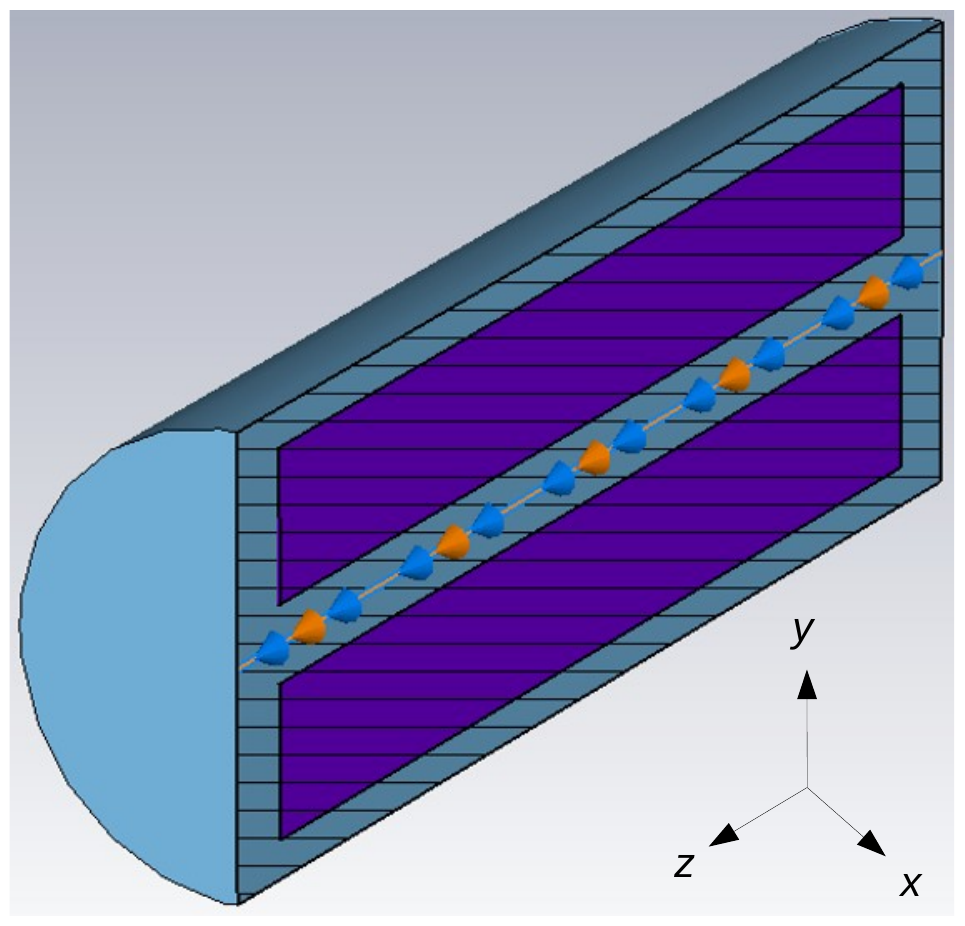}
\caption{Simulated collimator structure in CST. The beam simulated to generate and probe the wake potentials has the same properties of the witness bunch so far described.}
\label{CSTcollimator}
\end{figure}

\begin{figure}[h]
\centering
\includegraphics[width=1.0\linewidth]{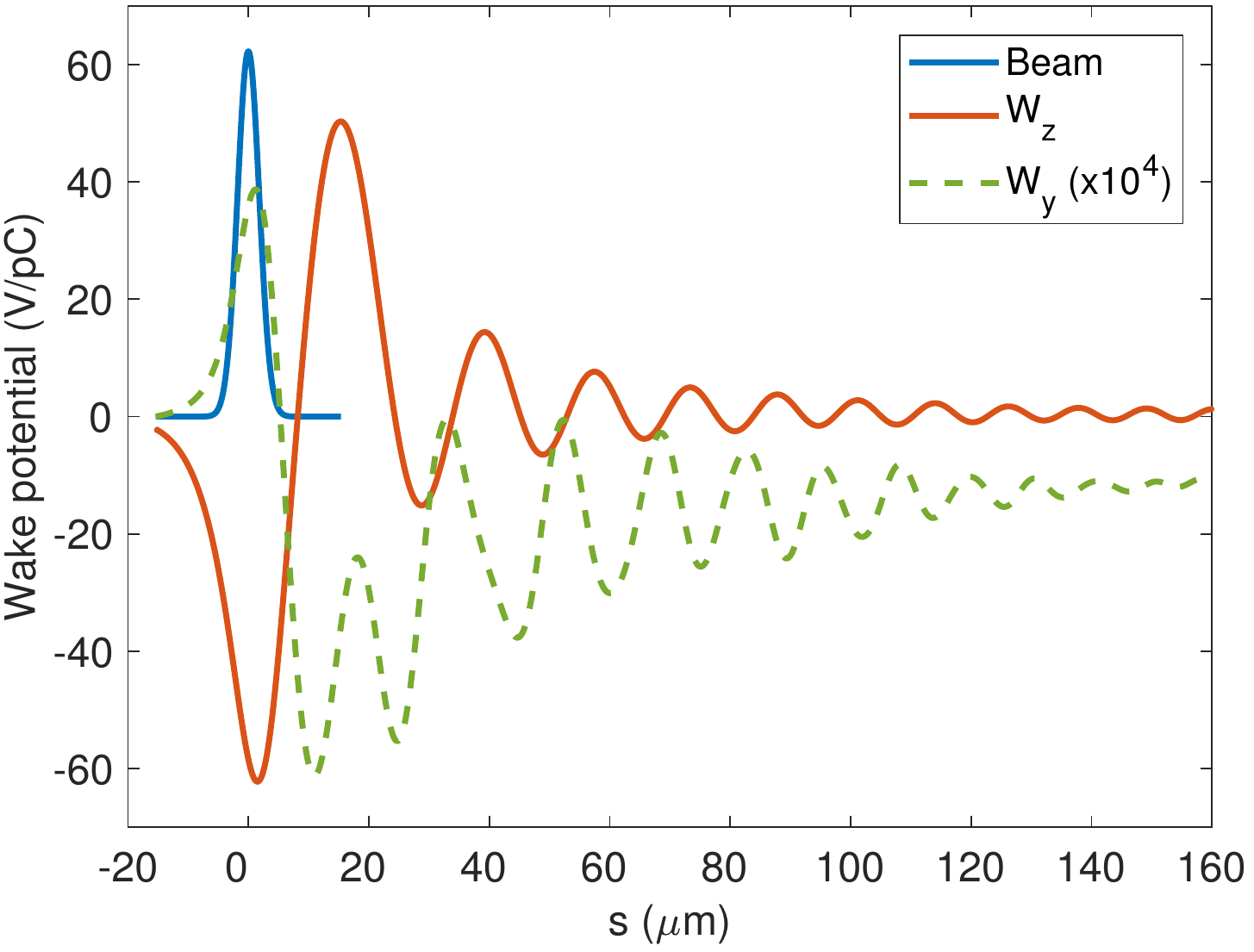}
\caption{Longitudinal (red) and transverse (green) wakefields induced into the collimator by the witness bunch (blue). To represent the lines with the same axis scale, the transverse wakefield $W_y$ (calculated at distance $r=50~\mu m$ from the collimator axis) has been enhanced by a factor $10^4$.}
\label{yzWakeCollimator}
\end{figure}

From the previous equations we thus have that the longitudinal wakes cause an energy decrease of the bunch while the transverse ones can affect the design orbit of the beam.
In order to quantify the effect of the wakes, we performed numerical 3D simulation with the CST code~\cite{cst_web}. Figure~\ref{CSTcollimator} shows the simulated collimator structure. The collimator aperture is $400~\mu m$ in diameter and 3~cm in length. The simulated beam, used to generate and probe at the same time the induced wake potentials, has the same properties of the witness bunch so far described. Both longitudinal $W_z$ and transverse $W_y$ wake potentials are calculated along the axis and with a $\Delta y=50~\mu m$ transverse offset, respectively. They are shown in Fig.~\ref{yzWakeCollimator} together with the witness bunch distribution.
From the simulations we obtained a longitudinal loss factor $k_l = 1.65$~kV/pC and a kick-factor $k_t=1.92$~V/pC/mm. From these loss parameters it is then possible to derive the energy decrease and the transverse kick received by the bunch. The energy loss can be easily computed as $\Delta E = k_l \cdot Q\approx 50$~keV. Similarly, the angular deflection produced by the transverse wake on a beam with energy $E$ can be estimated as~\cite{brynes2016studies} $\Delta\theta = \Delta y\cdot k_t\cdot Q/E\approx 2.88\times 10^{-9}$~rad.
From the resulting numbers we can thus conclude that the influence of the wakefields produced into the collimator by the witness bunch do not affect its dynamics and can be neglected.

\subsection{Particle-matter interactions}\label{geant_sim}
The interaction of the driver beam with the lead collimator has been numerically simulated by means of the  GEANT4 framweork, a single particle tracker which takes into account all the fundamental radiation-matter interactions~\cite{agostinelli2003geant4}.
GEANT4 is a toolkit for simulating the passage of particles through matter. It includes many functionalities like tracking, geometry, physics models and hits. Many physics processes are included and cover a wide range of interactions like electromagnetic, hadronic and optical processes, a large set of long-lived particles, materials and elements, over a wide energy range (from hundreds of eV up to TeV).
In defining and implementing all the involved components, all aspects of the simulation process have been included: (i) the geometry of the collimator system, (ii) the material involved (lead), (iii) the fundamental particles of interest (electrons), (iv) the generation of primary particles of events (electrons, hadrons and photons), (v) the tracking of particles through materials, (vi) the physics processes governing particle interactions (bremsstrahlung, pair-production, multiple-scattering, etc.).

For the current simulation, we proceeded as in the following. Firstly we transported with GPT the beam exiting from the first APL up to the collimator entrance. Here we imported and converted the GPT bunch in order to be treated with GEANT4. Finally the GEANT4 simulation output is imported again in GPT and used as input for the second APL.
In the GEANT4 simulation the FTFP\_BERT physics list has been adopted; it contains the standard electromagnetic and hadronic interactions, the latter ones implemented using the FTF parton string and Bertini cascade models~\cite{allison2016recent}. 

\begin{figure}[h]
\centering
\includegraphics[width=1.0\linewidth]{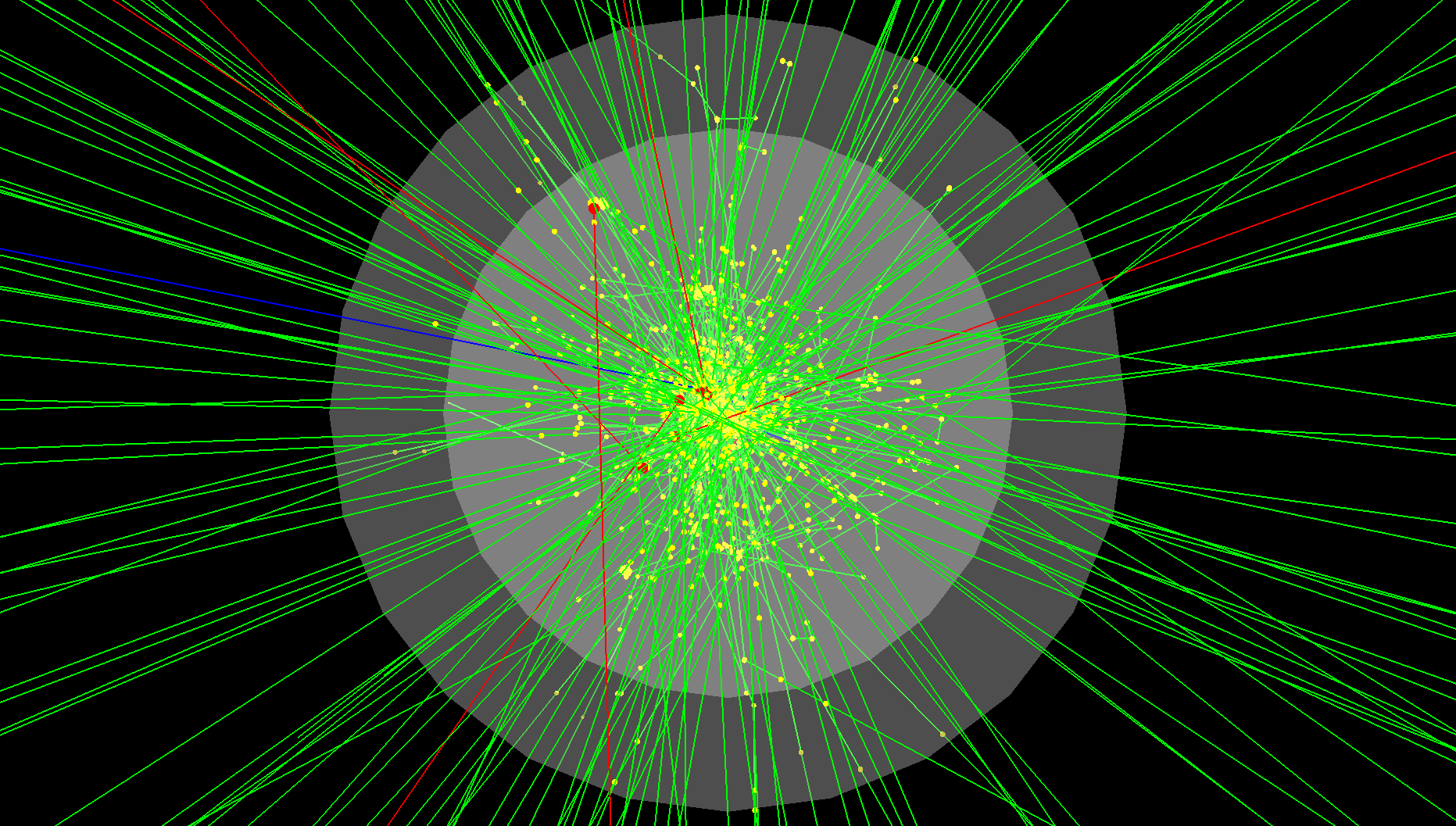}
\caption{Driver electron beam interacting with the collimator. The colors of the tracks stand for electrons (red), positrons (blue) and photons (green) in the $\gamma$ and X-rays range.}
\label{geant_driver}
\end{figure}

\begin{figure*}[t]
\centering
\subfigure{
\begin{overpic}[width=0.45\linewidth]{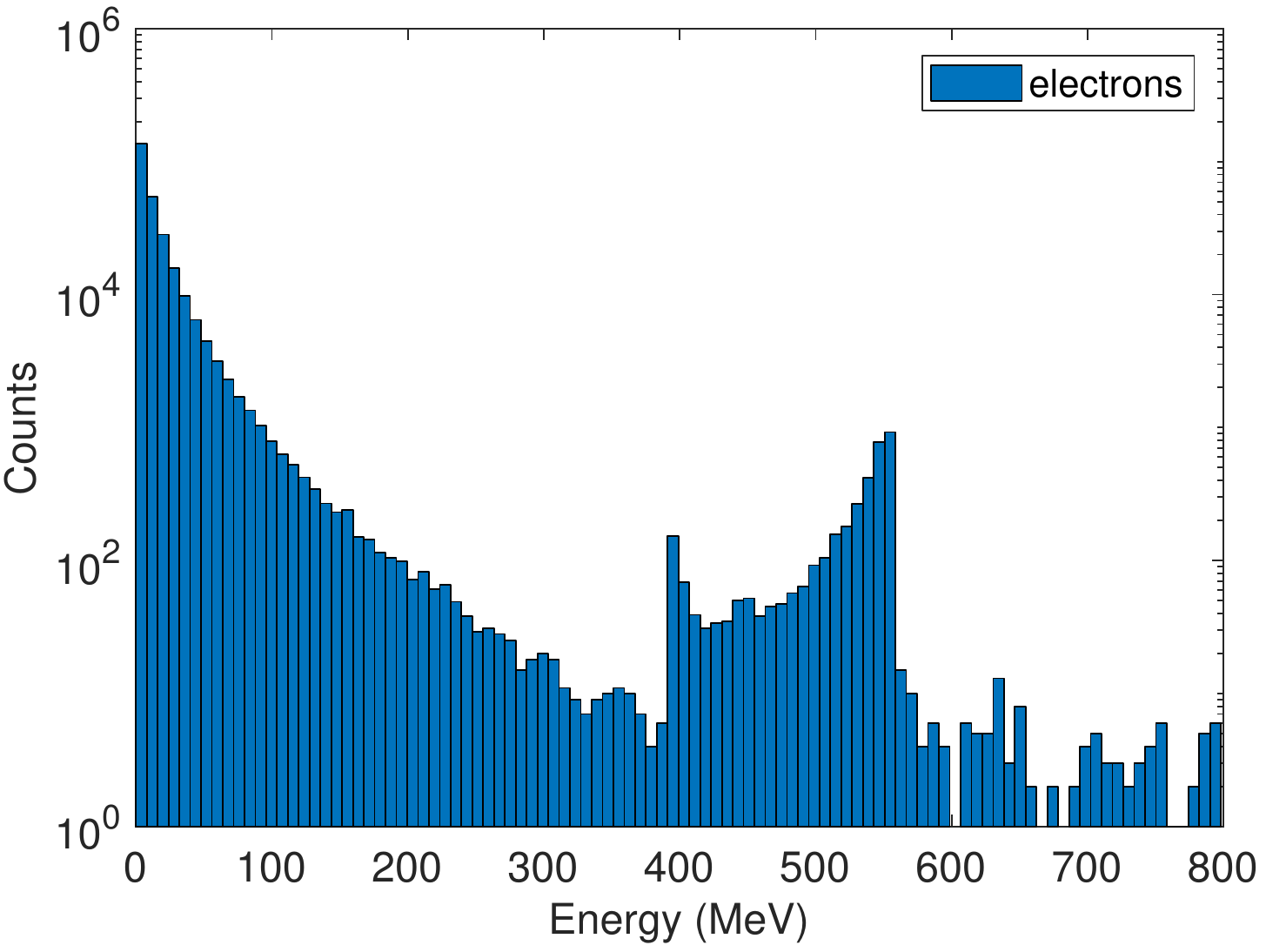}
\put(15,68){\color{black}\textbf{a}}
\end{overpic}
\label{GEANT_Electrons}
}
\subfigure{
\begin{overpic}[width=0.45\linewidth]{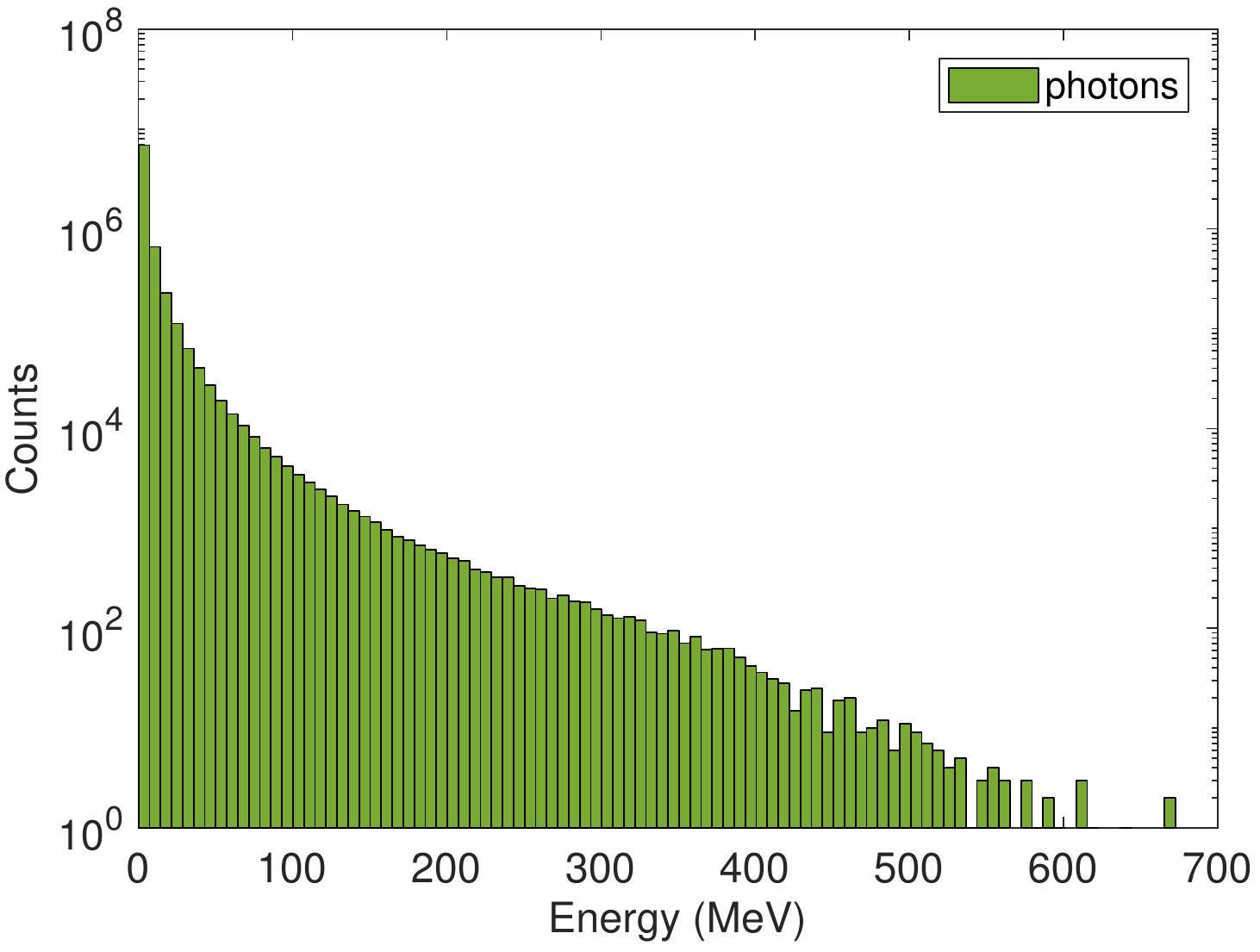}
\put(15,68){\color{black}\textbf{b}}
\end{overpic}
\label{GEANT_Photons}
}
\subfigure{
\begin{overpic}[width=0.45\linewidth]{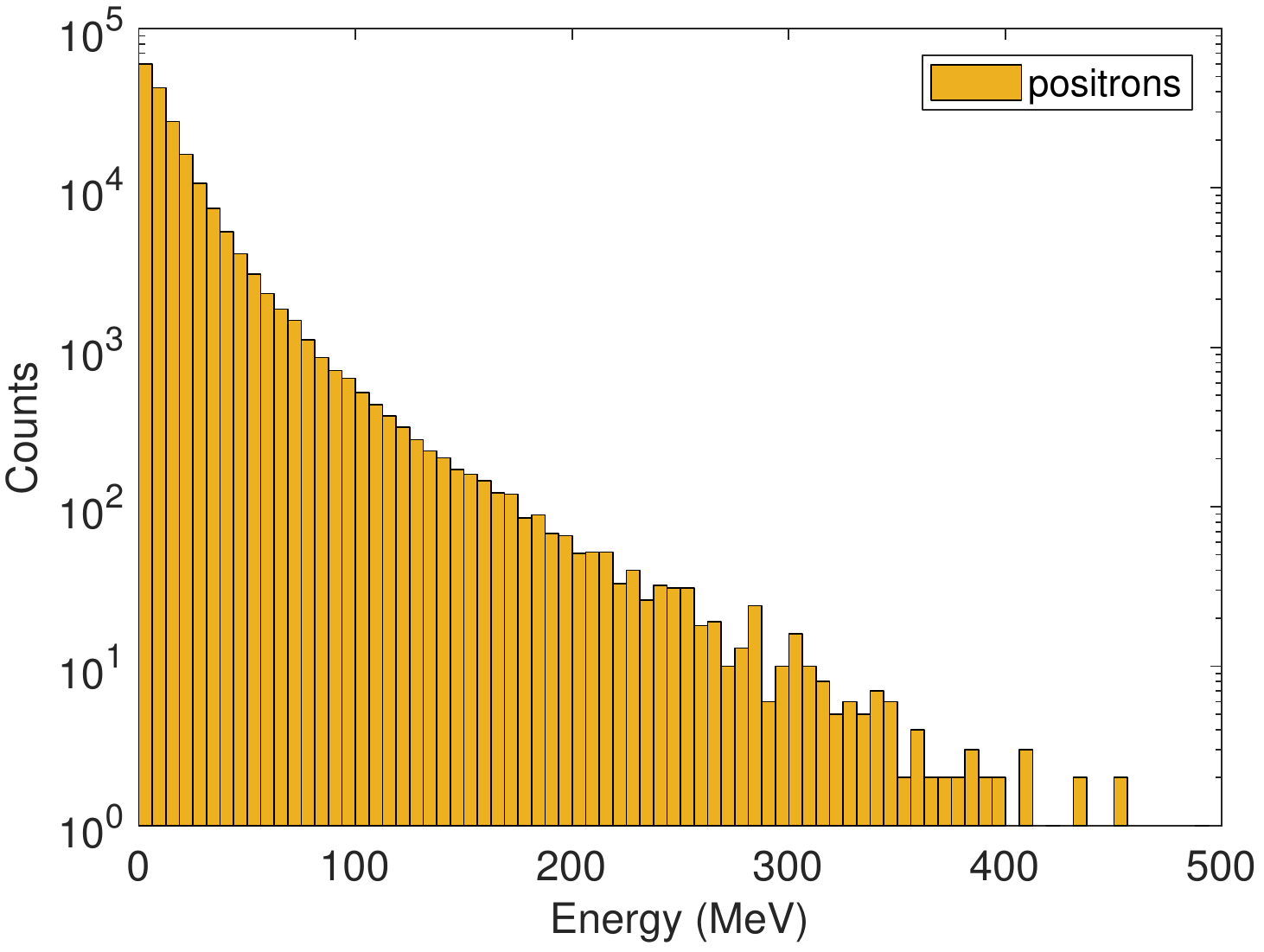}
\put(15,68){\color{black}\textbf{c}}
\end{overpic}
\label{GEANT_Positrons}
}
\subfigure{
\begin{overpic}[width=0.45\linewidth]{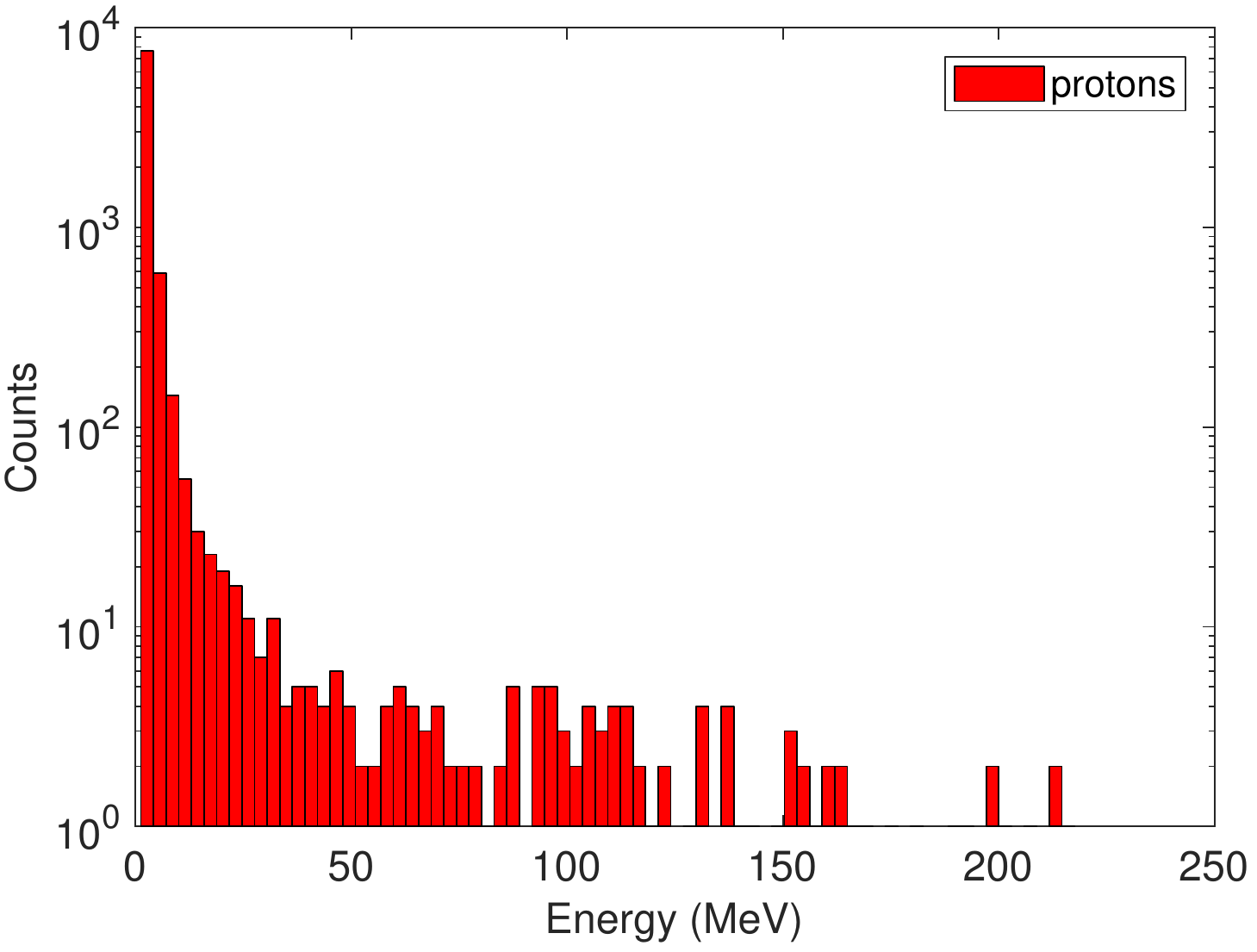}
\put(15,68){\color{black}\textbf{d}}
\end{overpic}
\label{GEANT_Protons}
}
\caption{Energy spectrum of the scattered and emitted electrons (a), photons (b), positrons (c) and protons (d) from the collimator after the interaction with the incoming electron beam. The particles have been collected over a $4\pi$ solid angle around the collimator.}
\label{geant_outpar}
\end{figure*}

The performances of the collimator have been optimized by varying four parameters in the simulation: the thickness, the inner and outer radii and the distance of the collimator from the first APL.
The final configuration reported in Tab.~\ref{system_properties} represents the best compromise regarding the driver dumping and the preservation of the witness beam charge. It consists of a 3~cm-long lead cilinder with outer radius of 1~cm and $200~\mu m$ radius aperture. According to the GEANT4 simulation, such a solutions allows to remove $\approx 98\%$ of the driver charge while the witness remains untouched.
Figure~\ref{geant_driver} shows the resulting tracks of the primary and secondary particles produced during the interaction of the driver bunch with the collimator.
The simulation recorded all the scattered and emitted particles after the interaction. In Fig.~\ref{geant_outpar} we show the counts of the electrons, photons, positrons and protons.
As shown, it results that a large amount of $\gamma$/X-rays are produced, up to $10^7$ by assuming $3\times 10^5$ incoming electrons. This is a potential source of background for any detector or diagnostics installed around the collimator that, thus, would require a proper shielding system to be adopted for any practical purpose.

\section{Conclusions}
The beam-driven Plasma Wakefield Acceleration technique represents one of the best candidates to develop next-generation compact accelerators. Being a new technology it must solve several issues in order to be adopted for any practical use as for user-oriented applications.
As highlighted in this work, a drawback is represented by the removal of the high-charge and energy-depleted driver bunch and, at the same time, the need to provide an efficient capture of the witness bunch avoiding its emittance degradation.
Here we have presented a possible solution, based on the use of active-plasma lenses, that has three main key features: (i) the tunability offered by the lenses themselves that allow to adapt the system to different configurations; (ii) a good efficiency in the driver bunch dumping, with only few percent of the incoming driver beam that remained at the end of the transport chain; (iii) the compactness of the entire solution, less than 1.4~m, in the treatment of GeV-class beams.
By looking to the scenario we have chosen as reference, i.e. the one represented by a proposed facility based on the EuPRAXIA design study, we have demonstrated that downstream the PWFA booster module, the accelerated 1~GeV witness can be efficiently captured and handled (in view of its transport up to the FEL undulators and user stations) with its normalized emittance that grew from $0.6~\mu m$ to $1.2~\mu m$ at the end of the extraction system. This number can be further optimized (and emittance better preserved) by implementing, for instance, plasma lenses with properly shaped density profiles~\cite{floettmann2014adiabatic,dornmair2015emittance}.
At the same time, we showed that a lead collimator located between the two plasma lenses is an affordable solution that allows to dump the 200~pC driver bunch to the level of few pC without affecting the witness. The study of the collimator, in particular, has been conducted by analyzing both the resistive wakefields excited along its aperture and the particle-matter interactions with the collimator walls.
The results confirmed that such a solution can be implemented in a future facility based on plasma acceleration where the compactness represents the mail goal.

\begin{acknowledgments}
This work has been partially supported by the EU Commission in the Seventh Framework Program, Grant Agreement 312453-EuCARD-2 and the European Union Horizon 2020 research and innovation program, Grant Agreement No. 653782 (EuPRAXIA).
\end{acknowledgments}

\bibliography{biblio}
\bibliographystyle{apsrev4-1}

\end{document}